# Spectral fingerprinting: Microstate readout via remanence ferromagnetic resonance in artificial spin ice

*Alex Vanstone*[*,1,2], *Jack C. Gartside*[1], *Kilian D. Stenning*[1], *Troy Dion*[1,2], *Daan M. Arroo*[2,3], *Will R. Branford*[1,4]

[1]Blackett Laboratory, Imperial College London, London, SW7 2AZ, United Kingdom. [2] London Centre for Nanotechnology, University College London, London, WC1H 0AH, United Kingdom. [3]Department of Materials, Imperial College London, London, SW7 2AZ, United Kingdom. [4] London Centre for Nanotechnology, Imperial College London, London, SW7 2AZ, United Kingdom.

*Corresponding author (av2813@ic.ac.uk)

# Abstract

Artificial spin ices are magnetic metamaterials comprising geometrically tiled interacting nanomagnets. There is significant interest in these systems for reconfigurable magnonics due to their vast microstate landscape. Studies to date have focused on the in-field GHz spin-wave response, convoluting effects from applied field, nanofabrication imperfections ('quenched disorder') and microstate dependent dipolar field landscapes. Here, we demonstrate the ability of zero-field artificial spin ice spectra to provide a 'spectral fingerprint' of the system microstate. Removing applied field allows deconvolution of distinct contributions to reversal dynamics from the spin-wave spectra, directly measuring dipolar field strength and quenched disorder. Mode amplitude provides population readout of differently magnetised nanomagnets, and hence net magnetisation measurement. We demonstrate microstate fingerprinting via distinct spectral readout of three microstates with identical (zero) magnetisation. These results establish zero-field 'spectral fingerprinting' as a rapid, scalable on-chip readout of both magnetic state and nanoscale dipolar field texture, a critical step in realising functional magnonic devices.

Artificial spin ice (ASI) are arrays of nanopatterned ferromagnetic arrays with frustrated inter-island dipolar interactions, leading to vastly degenerate low energy states. ASI systems were first intended as model systems mimicking magnetic frustration in rare-earth pyrochlores[1]. Scaling up atomic spins to 0.1-1 µm nanoislands allows system microstate readout using imaging techniques including magnetic force microscopy (MFM)[2]. Recently, ASI has found applications in novel computation[3–8] and reconfigurable magnonics[9,10].

Ferromagnetic resonance (FMR) spectroscopy measures spin-wave spectra and has proved a potent tool for studying ASI based reconfigurable magnonic crystals[11–17]. In a seminal work, Gliga *et al.*[11] predicted that FMR may be used for the quantitative detection of the population and separation of magnetic charge defects in square ASI. There is significant interest in using spin-wave spectra to identify ASI microstates[11,18–20], building on these proposals of measuring the presence and length of Dirac strings in square ASI[11]. A majority of ASI spin-wave studies focus on the in-field spectra, with only a few examples of measuring specific prepared states[17,19]. Resonant mode frequencies are a function of the external, demagnetisation, and local dipolar fields, providing rich information via the spectral response to field[21–23]. A limitation of in-field FMR is that varying applied field $H_{ext}$ changes the spectra in multiple different ways. When $H_{ext} \sim H_C$ (the array coercive field), the microstate evolves during reversal. In this field range the resonance frequency of any given mode changes due to increasing external field, changes in the demagnetizing field if the active island reverses and changes in the dipolar field landscape as neighbouring islands reverse. The precise microstate imprints subtle, informative details on the spin-wave spectra. However, these spectral shifts are dwarfed by mode frequency jumps associated with island reversal (~2 GHz vs 0.2 GHz)[17], limiting the microstate information revealed by field swept FMR.

Here, we employ zero-field FMR as a direct 'spectral fingerprint' readout of the microstate and nanoscale dipolar field texture. Removing the presence of external bias field, we can access fine microstate details across three ASI samples and deconvolute contributions to reversal dynamics arising from inter-island dipolar interaction and the Gaussian distribution of coercive fields[24,25] arising from nanofabrication imperfections termed quenched disorder. We extract absolute dipolar field magnitudes at specific lattice sites, challenging via alternative means. To illustrate the power of spectral fingerprinting and the additional information revealed versus magnetisation measurement, we prepare three distinct microstates with identical (zero) net magnetisation and observe starkly different spectra, giving rich microstate structure insight. These experiments demonstrate remanence FMR as a readout of both the microstate and nanoscale dipolar field texture at the nanoscale. This technique is widely applicable across a range of nanomagnetic systems[26–28] particularly in 3D artificial spin systems where directly reading microstates is challenging[29–31].

# Results and Discussion

## Remanence FMR

Remanence FMR functions by applying a microstate preparation field $H_{prep}$ then removing it and measuring zero field spectra. In zero external field $H_{ext} = 0$ the Kittel equation gives a nanoisland resonant frequency $f_0$:

$$f_0 = \mu_0 \gamma \sqrt{\left((H_{loc}) + (N_z - N_{H_\parallel}) \cdot M_s\right) \cdot \left((H_{loc}) + (N_{H_\perp} - N_{H_\parallel}) \cdot M_s\right)}$$

where $H_{loc}$ is the dipolar field from the surrounding bars, $N_{H_\parallel}$, $N_{H_\perp}$ the local demagnetisation factors along and perpendicular to the field respectively, $N_z$ the out-of-plane demagnetisation factor, $\gamma$ the gyromagnetic ratio (29.5GHz/T for permalloy), $\mu_0$ the magnetic permeability of free space, and $M_s$ the saturation magnetisation.

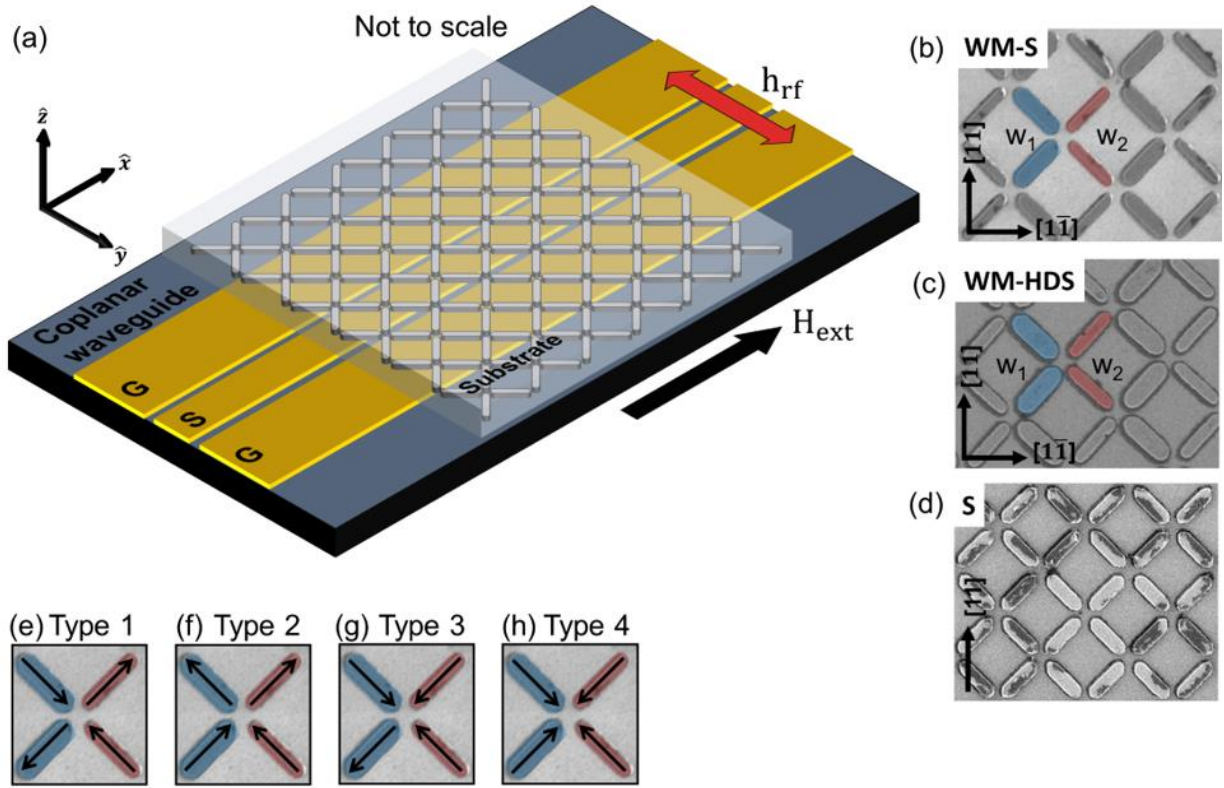

**Figure 1: Experimental measurement schematic and samples description** (a) Schematic of flip-chip FMR measurement. The sample is mounted on the coplanar waveguide with the bar's long axis at 45 degrees to the external field $H_{ext}$ which is perpendicular to the microwave field $h_{rf}$ generated by the waveguide. Dimensions of ASI to waveguide not to scale. (b, c) SEM of WM-S and WM-HDS samples. WM-S bars are 830 nm × 230 nm (wide-bar $w_1$) / 145 nm (thin-bar $w_2$) × 20 nm with 120 nm vertex gap (bar-end to vertex-centre). WM-HDS bars are 600 nm × 200 nm (wide-bar $w_1$) / 125 nm (thin-bar $w_2$) 20 nm with 100 nm vertex gap. Where the ground state (GS) is accessible via mounting the sample along the [11] crystallographic direction and monopole state (MS) is accessible via mounting the sample along the [1$\bar{1}$] direction. (d) SEM of the symmetric square, S sample, bar dimensions 474 × 135 × 20 $nm^3$, vertex gap 100 nm. It is mounted at 45 degrees to the bar's long axis or the [11] crystallographic direction. (e-h) Schematics of all four

vertex types for a square lattice. Types 1,2 are prepared by mounting the sample along the [11] direction and Types 2, 3, 4 are prepared by mounting the sample along the [1$\bar{1}$] direction

We consider three ASI samples, width-modified square (WM-S), figure 1b, width-modified high-density square (WM-HDS), figure 1c, and square (S), figure 1d comprising of identical bars. Width-modified samples allow for global field preparation of all four distinct vertex types (fig. 1e-h), see supplementary figure 1 for MFM images of pure microstates[32]. These samples allow us to demonstrate the spectral correspondence of remanence FMR when the microstate is well known, before using a range of disordered states in the conventional square ASI (S sample) as a proving ground for spectral microstate fingerprinting.

## Microstate control via Width-modification

WM-S and WM-HDS samples, shown in figures 1b, c are square ASI with sublattices of wide ($w_1$) and thin ($w_2$) bars. Wider bars have lower coercivity ($H_{c1} < H_{c2}$), allowing microstate control via the application of global field[32].

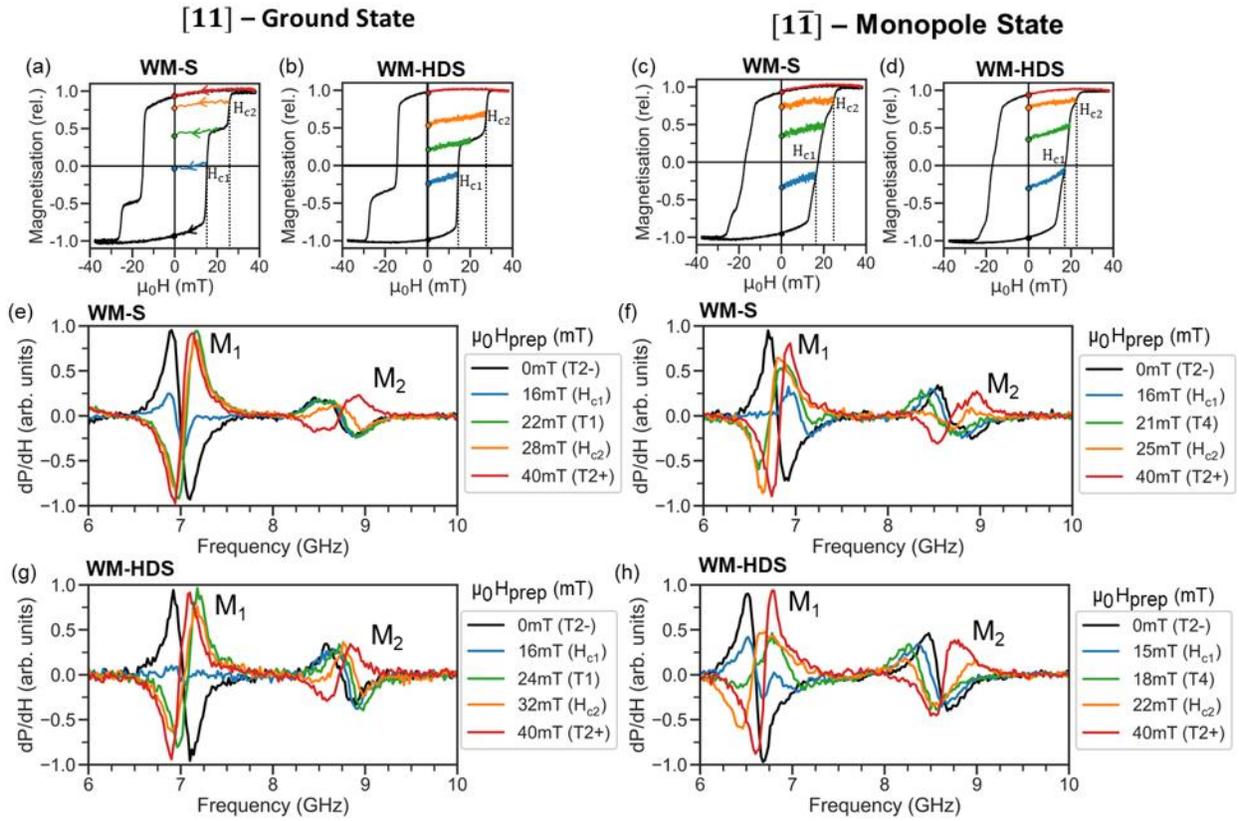

**Figure 2: Microstate control via width-modification and corresponding remanent spectra** (a-d) MOKE hysteresis loops in the ground state (GS) orientation for the WM-S (a) and the WM-HDS (b) sample. Hysteresis-loops in the monopole state (MS) orientation for the WM-S (c) and the WM-HDS (d) sample. Colour coded remanence magnetisation curves from the same preparation field as the spectra in (e-h) are also shown. (e-h) Differential remanence-FMR spectra for WM-S sample in GS (e) and MS (f) orientations and WM-HDS sample in GS (g) and MS (h) orientations. GS-orientation spectra corresponding to preparation fields for the Type 2- (black), $H_{c1}$ (blue), Type 1 (green), $H_{c2}$ (orange), Type 2+ (red) microstates. In the MS orientation the spectra shown correspond to the T2- (black), $H_{c1}$ (blue), Type 3/4 (green), $H_{c2}$ (orange), T2+ (red) microstates.

Mounting the width-modified samples along the crystallographic [11] axis and starting from a field saturated state type 2 (fig. 1f) prepares a type 1 microstate (fig. 1g) by applying a field such that only wide bars reverse. This is the system ground state[33–35] due to maximum dipolar flux closure, hence we term this field axis the 'ground state' (GS) orientation. Rotating the samples and applying $H_{ext}$ along [1$\bar{1}$] direction (fig. 1a, b) and reversing wide bars prepares the type 4 state (fig. 1h), with 4 like polarity magnetic charges at each vertex. This results in highly unfavourable dipolar field interactions, termed the 'monopole state' [36–38]. We hence term this sample mounting the 'monopole state' (MS) orientation. It also prepares the Type 3 state (fig. 1g) with 3 like polarity and 1 opposite polarity vertex charges if any angular misalignment is present.

MOKE hysteresis loops for WM samples are shown in figure 2a-d. Panels 2a,b show the ground state orientation WM-S (a) and WM-HDS (b) loops. Panels 2c,d show the monopole orientation WM-S (c) and WM-HDS (d) loops. The shape of the hysteresis loop changes significantly between the two orientations. The GS loops show two sharp steps in magnetisation at fields $H_{c1}$ and $H_{c2}$ corresponding to wide and thin bar reversal respectively. The magnetisation plateau between $H_{c1}$ and $H_{c2}$ corresponds to the type 1 state. MS hysteresis loops show gradual magnetisation reversal due to the energetically unfavourable dipolar field landscapes of type 3 and type 4 states. Here a clear plateau is absent, with kinks in the magnetisation curve revealing locations of the type 4 state in the WM-S sample (21mT) and type 3 state in the WM-HDS (20mT). The dipolar interaction in the WM-HDS sample is too strong for a pure type 4 state to be observed.

The remanence traces on the MOKE loops (coloured lines) leading from the major loop back to zero field show the resultant magnetization after the microstate preparation protocols. Figures 1e-h show remanence FMR spectra corresponding to microstates prepared by the remanence traces shown in figure 1a-d. The spectral plots show two major absorption peaks at ~ 7 GHz and ~

8.6 GHz corresponding to bulk centre-localised modes in wide ($M_1$) and thin ($M_2$) bars respectively. Spectra shown in figures 1e-h exhibit characteristic changes in the peak profiles of $M_1$ and $M_2$, both in terms of resonant frequency $f_0$ and differential amplitude $\frac{\partial P}{\partial H}$ depending on the microstate.

## Remanence FMR of Width-modified ASI

**Ground State Orientation Remanence FMR**

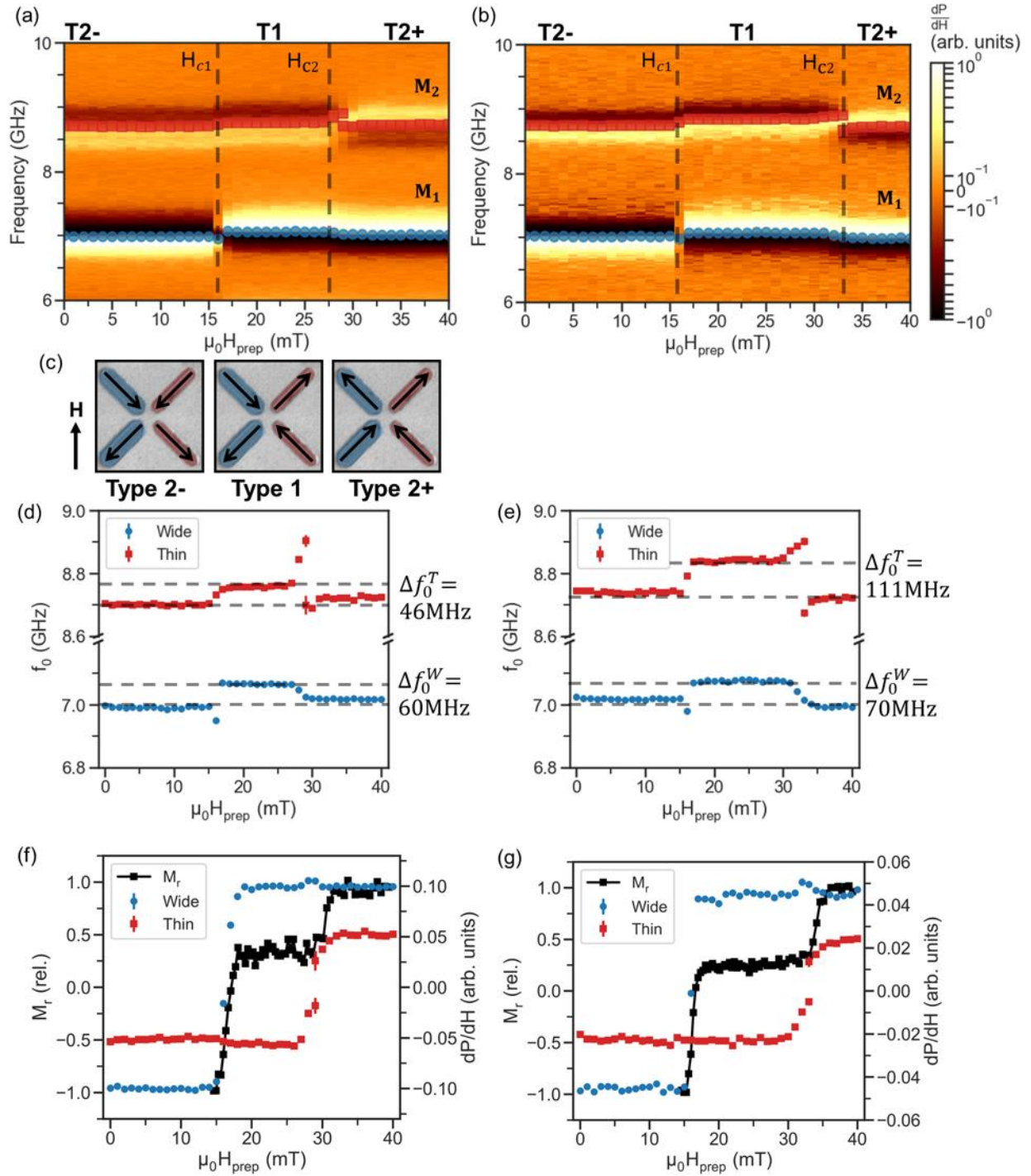

**Figure 3: GS orientation Remanence FMR and microstate evolution** (a, b) Heatmap of differential FMR amplitude as a function of frequency at a range of preparation fields H_prep in the GS orientation for WM-S (a) and WM-HDS (b) samples. Magnetisation schematics of the type 2-/type 1/type 2+ microstate trajectory are shown in (c). (d, e) Extracted resonant frequencies $f_0$ for the bar-centre localised modes are shown on scatter plots in blue (wide) and red (thin) for WM-S (d) and WM-HDS (e). Samples were initially saturated in -200 mT before a preparation field $H_{prep}$ was applied in a positive direction and the FMR spectra were measured in zero field. This was repeated for $H_{prep}$ in 1 mT steps from 0-40 mT. Coercive fields are marked $H_{c1}$ (wide bar) and $H_{c2}$ (thin), accompanied by a phase reversal of the relevant mode in the differential plot. (f, g) MOKE measured remanence magnetisation (black trace, left y-axis) and FMR amplitude of the $M_1$ (blue trace, right y-axis) and $M_2$ (red trace, right y-axis) modes for WM-S (f) and WM-HDS (g).

With $H_{ext}$ along the [11] orientation, samples were saturated into the type 2- state by applying an initial -200 mT field. The preparation field $H_{prep}$ was then stepped from 0-40 mT in 1 mT steps, measuring zero field spectra after each field. Figures 3a,b show differential spectral heatmaps of the preparation field against FMR frequency. When wide bars reverse at $H_{c1}$, the sample switches from the type 2- to the type 1 state. Thin bars reverse at $H_{c2}$, switching from the type 1 to the type 2+ state. Due to the stable energetics of the type 1 ground state, $H_{c2}$ is increased in this configuration relative to an isolated thin bar, broadening the type 1 field window.

As the sample enters a type 1 state the local dipolar field landscape shifts, blueshifting both modes in figures 3d,e. $M_1$ blueshifts by $60 \pm 7$ MHz (WM-S) and $70 \pm 5$ MHz (WM-HDS) while $M_2$ blueshifts $46 \pm 8$ MHz (WM-S) and $111 \pm 8$ MHz (WM-HDS). Due to the smaller lattice parameter and larger dipolar field of WM-HDS, frequency shifts are enhanced relative to the WM-S sample. Similarly, this explains the relative difference in shift magnitude between $M_1$ and $M_2$.

The dipolar field emanating from the wide bar is stronger due to its larger volume, so it induces a greater frequency shift on the thin bar, while the frequency shift in the $M_1$ mode is smaller.

Fitting the in-field FMR response with the Kittel equation to $M_1$ and $M_2$ modes, we extract the relative shift in the dipolar field from the $f_0$ shift when the microstate changes from type 2 to type 1. For the WM-S (WM-HDS) samples, this gives $1.9 \pm 0.1$ mT ($4.6 \pm 0.1$ mT) for the thin bar and $2.5 \pm 0.1$ mT ($2.9 \pm 0.1$ mT) for the wide bar. Point dipole simulations estimate type 2 to type 1 dipolar field shifts as 1.9 mT (4.6 mT) and 4.5 mT (7.2 mT) at the centre point of the thin and wide bars respectively for WM-S (WM-HDS) samples. The WM-S sample demonstrating efficacy of spectral fingerprinting at providing an absolute measurement of dipolar field textures. The discrepancy in the WM-HDS may arise from significant edge-curling, suggesting that the use of the field at the island centre instead of integrating over the whole mode area is not a good approximation for strongly-interacting samples.

The extracted amplitude of $M_1$ and $M_2$ are shown in figures 3c,d along with remanence magnetisation $M_r$ for WM-S and WM-HDS samples. Remanence magnetisation was measured via MOKE using different $H_{prep}$ and taking remanence curves, figures 2a-d. $H_{c1}$ and $H_{c2}$ match with the sign change of remanence FMR measurements, allowing effective magnetisation measurement of each bar subset.

**Monopole State Orientation Remanence FMR**

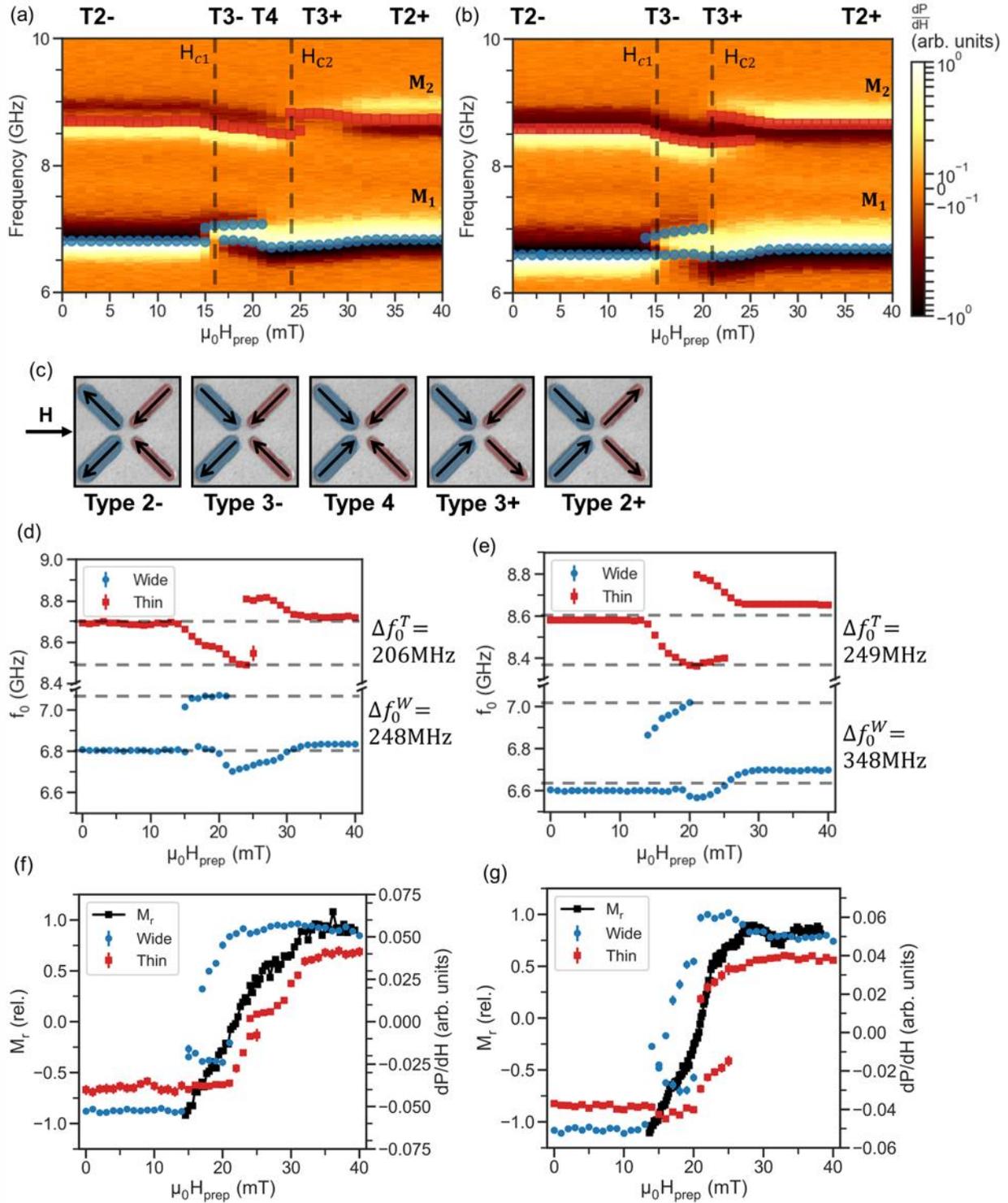

**Figure 4: MS orientation Remanence FMR and microstate evolution** (a, b) Heatmap of differential FMR amplitude vs frequency at a range of preparation fields $H_{prep}$ in the MS

orientation for WM-S (a) and WM-HDS (b) samples. Magnetisation schematics of the type 2-/type 3-/type 4/type 3+/type 2+ microstate trajectory are shown in (c). (d, e) Extracted resonant frequencies $f_0$ for the bar-centre localised modes are shown on scatter plots in blue (wide bar) and red (thin bar) for WM-S (d) and WM-HDS (e). Samples were initially saturated in -200 mT before sweeping $H_{prep}$ 0-40 mT in 1 mT steps and measuring zero field FMR spectra between each step. Coercive fields are marked $H_{c1}$ (wide) and $H_{c2}$ (thin), accompanied by a phase reversal of the relevant mode in the differential plot. (f, g) MOKE measured remanence magnetisation (black trace, left y-axis) and FMR amplitude of wide bar (blue trace, right y-axis) and thin bar (red trace, right y-axis) modes for WM-S (f) and WM-HDS (g).

Rotating the sample along the MS orientation, samples are saturated in type 2- state at -200 mT and zero-field FMR spectra are again measured after applying preparation fields from 0-40mT with 1mT steps. Figure 4a,b shows spectral heatmaps of the preparation field against FMR frequency. Two modes are observed corresponding to wide (lower frequency) and thin bars (higher frequency). Type 4 vertices are energetically unfavourable hence transition in and out of this state occurs via the type 3 state (fig. 4c). The field window for pure type 4 microstates is further reduced by the dipolar field from reversed wide bars lowering $H_{c2}$. Like the hysteresis loops shown in figure 1c, d, the MS-orientation heatmaps show a continuous frequency shift in the $M_1$ and $M_2$.

As the population of type 3 and type 4 vertices grows the local dipolar field landscape changes, shifting the resonant frequency of both modes figure 4c,d. Due to quenched disorder, type 3+ vertices populate the sample gradually. Type 3+ vertices increase $M_1$-mode splits as the reversed wide bars blueshift and the $M_2$ mode gradually redshifts. Once $H_{prep} > H_{c1}$ the population of type 4 vertices increases resulting in greater blueshift in $M_1$ and redshift in $M_2$ until the maximum type 4 population is reached. For the WM-S (WM-HDS) samples total $M_1$ blueshift is 248 ± 4 MHz (348 ±19 MHZ) while the $M_2$ redshifts 206 ± 15 MHz (249 ± 7 MHz) in WM-S (WM-HDS). In the MS orientation the resonance frequency continuously shifts due to gradually changing the average local dipolar field from the ensemble of microstates throughout the reversal process. The mix of vertex population and quenched disorder leads to a more disordered transition into and out of type 4 and consequently creates a more varied dipolar field landscape.

Using Kittel fits, the relative MS orientation shift in dipolar field between type 2 and type 4 states is 8.6 ± 0.2 mT (10.6 ± 0.1 mT) and 10.5 ± 0.1 mT (15.0 ± 0.3 mT) in WM-S (WM-HDS) for thin and wide bars respectively. Point dipole modelling estimates bar centre-point shifts of 7.2 mT (11.6 mT) and 4.5 mT (7.3 mT) for WM-S (WM-HDS) in thin and wide bars respectively. The simulated dipolar field estimate underestimates the dipolar field shift in the wide bars again due to neglecting realistic magnetization textures such as edge curling influence on the mode shift.

Extracted $M_1$ and $M_2$ amplitudes are shown in figure 3e,f along $M_r$ for WM-S and WM-HDS samples. The MS orientation remanence magnetisation curve does not show a distinct magnetisation plateau due to the similar coercive fields of the thin and wide bars in this direction. In FMR measurements the distinct resonance frequency of the thin and wide bars allows each subset to be probed individually, in contrast to conventional magnetometry which can only access the bulk magnetisation. The FMR differential amplitude of $M_1$ and $M_2$ correspond well with the remanence magnetisation measurement.

# Remanence FMR of Symmetric Square ASI

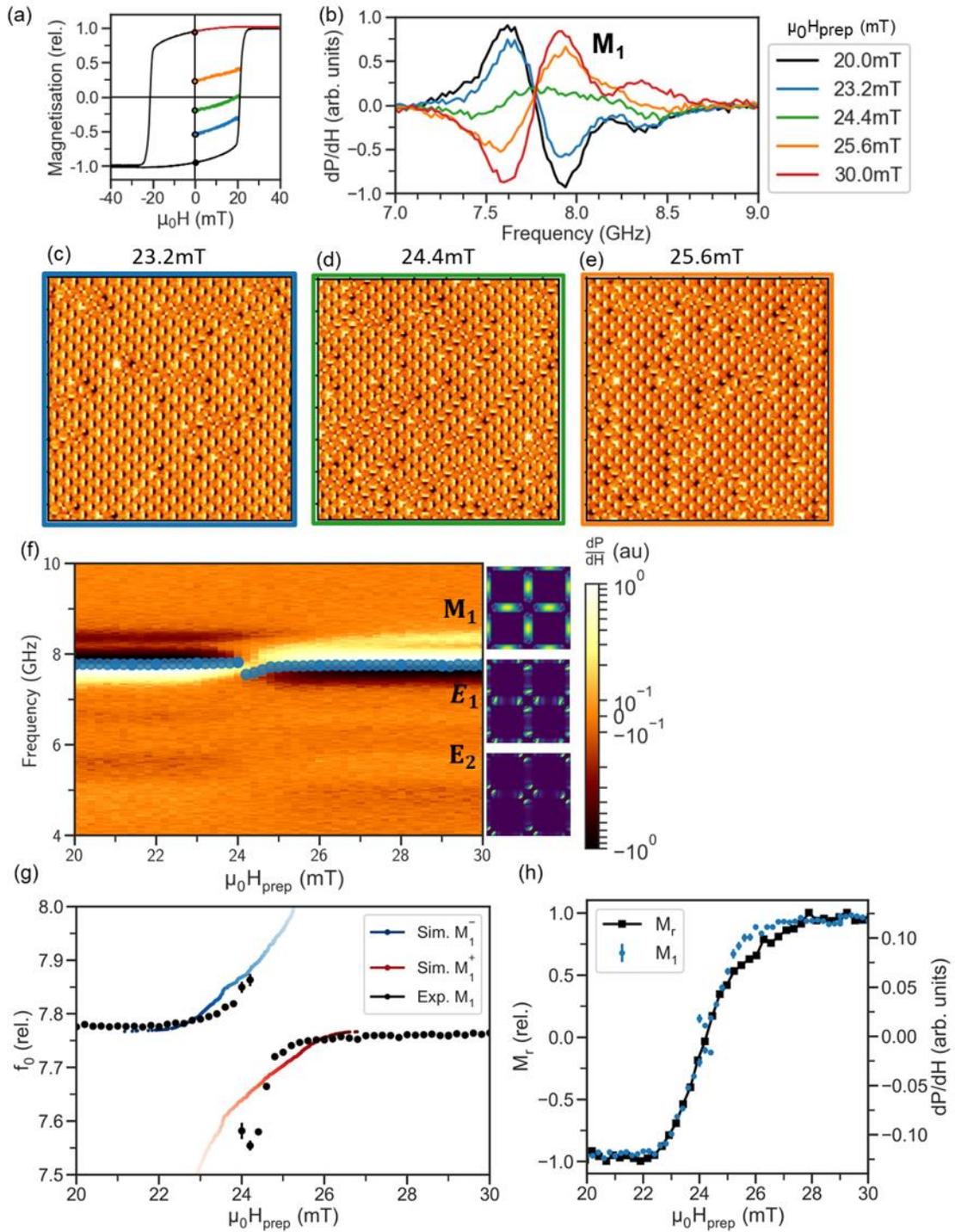

**Figure 5: Remanence FMR and microstate evolution of symmetric square ASI** (a) MOKE Hysteresis-loop of S sample showing remanence traces at $H_{prep} = 20, 23.2, 24.4, 25.6$ mT and 30

mT. (b) Remanence FMR spectra for microstates corresponding to $H_{prep}$ values from MOKE and MFM measurements. (c-e) MFM images of microstates corresponding to the same intermediate $H_{prep}$ values are shown on MOKE loop and FMR spectra. (f) Remanence FMR heatmap for $H_{prep}$ = 20-30 mT, 0.2 mT field steps, 20 MHz freq. steps with normalised spectral powermaps of centre-localised bulk mode $M_1$, edge-localised bulk mode $E_1$ and edge mode $E_2$. (g) Fitted resonance frequency of $M_1$ mode (black) as a function of preparation field $H_{prep}$ with point dipole modelled resonance frequency shift of the positive $M_1^+$ (blue) and negative $M_1^-$ (red) modes, where the transparency corresponds to the amplitude of the mode. (h) Remanent MOKE magnetisation (black, left y-axis) and FMR power (blue, right y-axis) as a function of $H_{prep}$.

We now apply spectral fingerprinting to symmetric square ASI (S sample) comprising of identical bars. Where width-modified samples follow a well-defined microstate trajectory during reversal, the S-S sample evolves through disordered microstates with mode frequencies determined solely by local dipolar field texture and quenched disorder.

Figure 5a shows the MOKE hysteresis loop with highlighted remanence magnetisation traces at a range of $H_{prep}$ and corresponding to the prepared microstates identified via MFM images (fig. 5c-e) and differential FMR spectra (fig. 5b). The sample was saturated along -x before measuring MFM and FMR. Spectra exhibit two dominant modes corresponding to unreversed $M_1^-$, (-x magnetised, 20 mT trace) and reversed $M_1^+$ (+x magnetised, 30 mT trace) bars, with partially reversed microstates a combination of both *(i.e.* 24.4 mT trace). Figure 5f shows remanence spectral heatmaps with $H_{prep}$ = 20-30mT, 0.2 mT steps. We observe 2 additional lower intensity modes in addition to the dominant bulk centre-localised mode $M_1$, corresponding to bulk edge-localised $E_1$ and the edge $E_2$ modes with simulated spatial powermaps for each mode. All modes exhibit characteristic sign change (fig. 5g) of $\frac{\partial f}{\partial H}$ around $H_c$ associated with magnetisation reversal.

At remanence, isolated reversed and unreversed bars will have the same resonant frequency. However, in a lattice the surrounding bars and lattice imperfections lead to resonant frequency shifts throughout reversal. Within the coercive field distribution both $M_1^-$ and $M_1^+$ contribute and are identified via careful fitting of two opposite sign differential Lorentzian curves. The extracted $f_0$ figure 5f shows a characteristic asymmetric frequency shift around $H_c$ from the combination of local dipolar field texture and quenched disorder.

The distribution of coercive fields over the sample follows a distribution of bar dimensions due to imperfections in the nanofabrication (quenched disorder) which varies the demagnetisation factors and hence $f_0$. This can be interpreted as a distribution of widths with on average wider bars having lower $H_c$ and lower $f_0$. These bars will typically reverse at lower field and upon reversal will be surrounded by unreversed neighbouring bars, experiencing local dipolar field oriented opposite their magnetisation. This reduces $H_{eff}$ and hence redshifts $f_0$. The total frequency shift for reversed bars is a sum of the redshift due to quenched disorder and local dipolar field $\Delta f_0^+ = -\Delta f_0^{QD} - \Delta f_0^{dip}$. The corollary of these arguments is that bars reversing at higher fields typically have higher shape anisotropy and hence higher $f_0$. The dipolar field frequency shift will still result in an $f_0$ redshift as the symmetry of the argument remains the same and so total frequency shift for unreversed bars is $\Delta f_0^- = \Delta f_0^{QD} - \Delta f_0^{dip}$. This asymmetry in the contribution of the dipolar field to frequency shift results in $M_1^-$ showing a smaller frequency shift than $M_1^+$ throughout reversal. The square reversal was modelled for the reversed (red) and unreversed (blue) bar population using point dipole simulations, including a quenched disorder of 4% and converting the $H_{eff}$ at each bar into $f_0$ via MuMax3 simulations. In figure 5g simulated curves generated via point dipole simulation are superimposed for the reversed (red) and unreversed (blue) bar populations, including quenched disorder of 4% and converting $H_{eff}$ at each bar to $f_0$ via results from MuMax3 simulation. Close correspondence is observed between measured and simulated behaviour, confirming the attribution of the initial low field increase in $f_0$ to quenched disorder and the later higher field sharp decrease in $f_0$ to the local dipolar field texture. Crucially, this allows absolute measurements of both. Deconvoluting these effects in strongly interacting magnetic nanoarrays is historically extremely challenging, and the demonstration here is a key strength of the spectral fingerprinting method.

Figure 5h shows a comparison of MOKE remanence magnetisation measurement (black curve, left y-axis) with $M_1$ FMR amplitude. As in figures 3,4 f-g, an extremely close correspondence is observed – illustrating the effectiveness of spectral fingerprinting at elucidating not just fine microstate details, but also the system magnetisation.

# Remanence FMR analysis of multiple M = 0 microstates

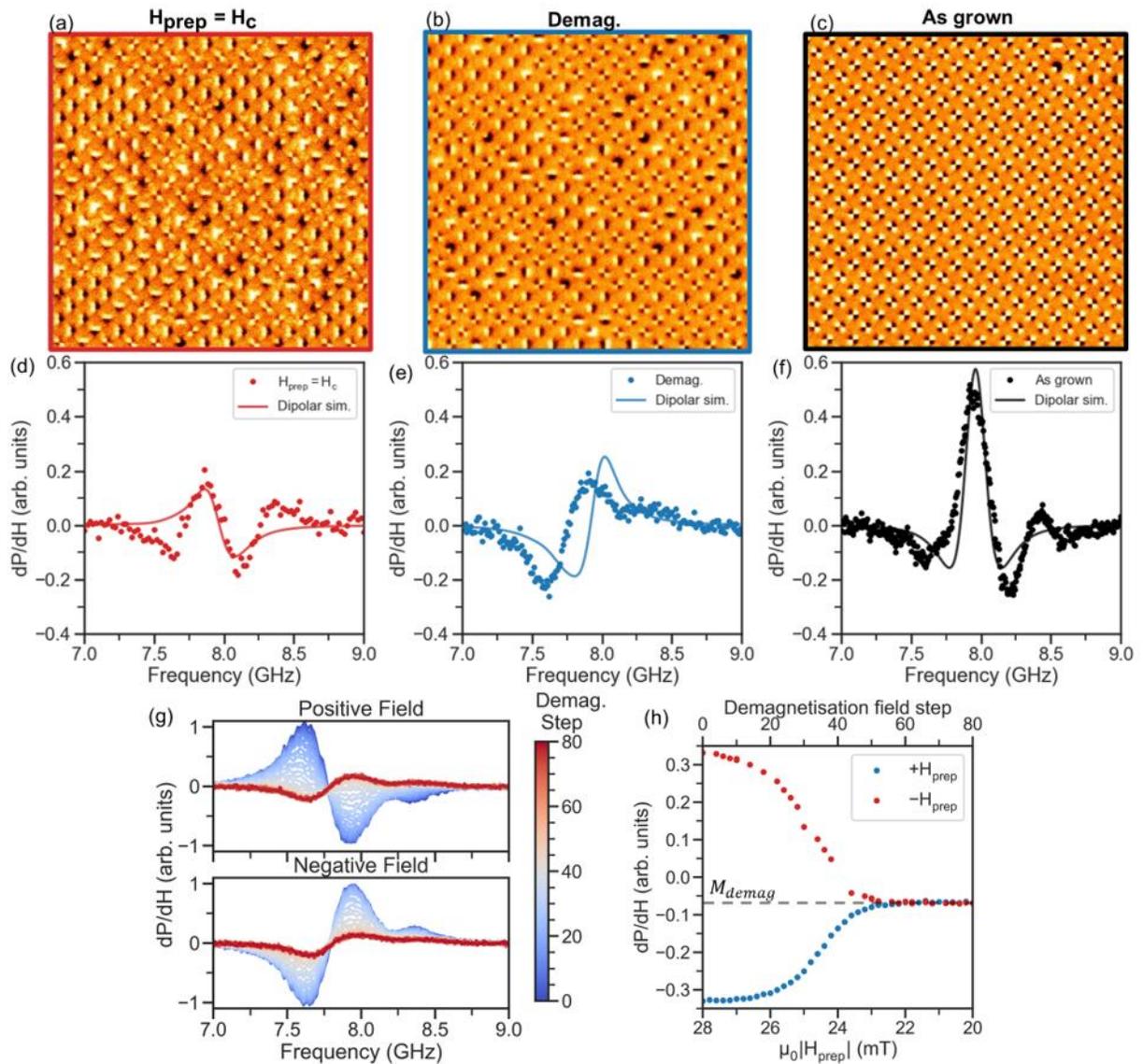

**Figure 6: Remanence-FMR of multiple M = 0 Microstates** (a-c) MFM images of the S sample taken in three M = 0 states. The as-grown state (a), AC-field demagnetised state (b) and a state prepared by applying positive field equal to $H_c$ after negative saturation (c). (d-f) Remanence-FMR spectra of the three M = 0 states shown in MFM images (a-c) as a scatter plot with dipolar simulations using the microstate shown above of the FMR spectra plotted as a line. (g) Remanence-FMR spectra in both positive and negative preparation-fields from 28 mT to 20 mT measured

during an AC field demagnetisation process with field steps of 0.1 mT. Colour bar indicated field magnitude applied before measuring remanence spectra. (h) FMR mode amplitude measured during AC field demagnetisation. FMR amplitude is presented both as a function of demagnetisation step number (top x-axis) and absolute preparation-field magnitude (bottom x-axis). The residual mode amplitude after demagnetisation due to remaining net magnetisation $M_{demag}$ is indicated by the dashed line.

Here we compare three M = 0 microstates; prepared by applying $H_{prep} = H_c$ after negative saturation (6a), AC field demagnetisation (6b), and as grown (6c). Macroscopic magnetisation measurements are unable to distinguish between these M = 0 states, but their remanence FMR spectra are distinct due to different local dipolar field textures in each microstate. The microstate in 6a is dominated by small domains of type 2 vertices magnetized in random directions, the microstate in 6b is a roughly equal mixture of type 1 and randomly oriented type 2 domains and the as grown sample is close to perfectly periodic type 1 order. Figure (6d-f) shows the remanence spectra for the three microstates. The MFM images were read into a dipolar simulation and FMR spectra generated by summing together differential Lorentzian peaks for each macrospin with resonant frequency estimated via the local dipolar field at each bar. The $H_c$ state spectra shows a low amplitude mode indicating a wide distribution of dipolar field landscapes with equal populations of negatively and positively magnetised bars, reflected in the low amplitude of the simulated curve.

The demagnetised state shows a shift in the resonance frequency because of of the high population of type 1 states but only a single magnetisation direction. The progression to the demagnetised state can be seen in figures 6g,h as remanence FMR spectra were measured throughout AC demagnetisation from 28mT- 20mT in 0.1mT steps. The sample starts in a saturated type 2 state and as the preparation field is reduced, bars with higher-than-average coercive field lock into the last magnetisation state where the field was large enough to reverse. As the $H_{prep}$ spectra in the positive field direction changes phase while in the negative field direction it remains the same and the net FMR power does not reach zero (fig. 6h). This is evidence of residual magnetisation $M_{demag}$ due to the imperfection of demagnetisation routines in reaching the ground state[39,40]. The different populations of type 1 and type 2 vertices results in the difference between the M=0 spectra.

In the as-grown state the vertices of square ASI will tend towards type 1 (fig. 6c) as the lowest energy configuration. The spectrum (fig. 6f) shows a superposition of $M_1^+$ and $M_1^-$ with a slight shift in frequency between the two modes, however, the antiferromagnetic ordering of the ground state results in an identical dipolar field on both the positively and negatively magnetised bars. In zero field this will result in a cancellation of the positively and negatively magnetised modes. However, incorporating a slight magnetic field of -0.5mT, possibly due to trapped flux, into the simulation reproduces the splitting of the $M_1^+$ and $M_1^-$ modes. Despite the small sample area of the MFM measurements the simulated and experimental spectra for the M=0 microstates agree very well.

# Conclusion

We have demonstrated 'spectral fingerprinting' across a range of strongly interacting nanomagnetic arrays. Via zero field FMR, spectral fingerprinting bridges the gap between bulk magnetisation measurements such as MOKE or VSM, and single macrospin resolution microstate mapping such as MFM or PEEM. Operating at a fraction of the time of single macrospin mapping and inherently scalable (demonstrated here on mm scale arrays), spectral fingerprinting provides information unavailable using net magnetisation measurements. Measuring microstate dependent absolute dipolar field magnitude has long been a goal of research into interacting nanomagnetic systems. Here we have provided an elegant solution requiring an off the shelf FMR system, with the underlying methodology equally applicable across alternative spin-wave measurements such as Brillouin Light Scattering[41,42]. While our demonstration here has concentrated on artificial spin ice, spectral fingerprinting is ideally suited across a range of interacting nanomagnetic systems[28] particularly the burgeoning range of 3D artificial spin systems where single macrospin imaging is inherently much harder[29,30]. Additionally, spectral fingerprinting is an attractive state readout solution for recent neuromorphic and wave computation schemes harnessing the vast set of microstate spaces for next-generation computing[5–7].

# Methods

## Simulation details

To determine the frequency shift as a result of quenched disorder and the local dipolar field we use a combination of experimental Kittel fitting, micromagnetic simulation and dipolar needle simulations. We fit the Kittel equation to the resonance frequency as a function of field using to experimentally to determine the remanence frequency shift as a result of the local dipolar field. The micromagnetic simulations use the same dimensions as the sample with simulation parameters $M_S$ of 700kA/m, α of 0.006 to determine the shift in frequency in as a result of the width of the sample. While the quenched disorder affects many properties in the sample, to create a Gaussian variation in the coercive field of the nanobars we assume that this can be modelled through changes in the width of the sample. We then use a point dipole model to simulate the local field of a 30x30 lattice with Gaussian spread of coercive fields of 4% to represent quenched disorder and an interaction strength of 1.08mT ($k = (\mu_0 M)/(4\pi a^3)$) where ($a = 287$ nm) is the lattice spacing, M is the absolute magnetic moment of a single magnet. We track the coercive field and local dipolar field on both the reversed and unreversed bars. The average coercive field is translated into a deviation in width and from the micromagnetic simulations we can estimate the shift in the resonance frequency as a result of the quenched disorder. The average local dipolar field is substituted into the Kittel equation to estimate the shift in the resonance frequency due to the dipolar field. Summing these two effects gives the shift in the resonance frequency as the S sample goes through reversal.

## Experimental details

Samples were fabricated via electron beam lithography lift-off method on a Raith eLine system with PMMA resist. $Ni_{81}Fe_{19}$ (permalloy) was thermally evaporated and capped with $Al_2O_3$. In the width-modified samples a 'staircase' subset of bars was increased in width to reduce its coercive field relative to the thin subset, allowing independent subset reversal via global field.

Ferromagnetic resonance spectra were measured using a NanOsc Instruments cryoFMR in a Quantum Design Physical Properties Measurement System. The measurements were all carried out at room temperature. Broadband FMR measurements were carried out on large area samples (2x2 mm$^2$) mounted flip-chip style on a coplanar waveguide at 45 degrees to the bar's axis, the [11] or [1$\bar{1}$] crystallographic directions. The waveguide was connected to a microwave generator, coupling RF magnetic fields to the sample. The output from waveguide was rectified using an RF-diode detector. Measurements were done in fixed in-plane field while the RF frequency was swept in 20 MHz steps. The DC field was then modulated at 490 Hz with a 0.48 mT RMS field and the diode voltage response measured via lock-in. The experimental spectra show the derivative output of the microwave signal as a function of field and frequency. The normalised differential spectra are displayed as false colour images with symmetric log colour scale.

Magnetic force micrographs were produced on a Dimension 3100 using commercially available normal moment MFM tips.

MOKE measurements were performed on a Durham Magneto Optics NanoMOKE system. The laser spot is approximately 20 μm diameter. The longitudinal Kerr signal was normalized, and the linear background subtracted from the saturated magnetisation. The applied field is a quasistatic sinusoidal field cycling at 11 Hz and the measured Kerr signal is averaged over 300 field loops to improve signal to noise.

ASSOCIATED CONTENT

**Supporting Information**.

The following files are available free of charge.

MFM images of prepared states Type 1-4 in width-modified sample (PDF)

AUTHOR INFORMATION

**Corresponding Author**

Alex Vanstone - Blackett Laboratory, Imperial College London, London SW7 2AZ, United Kingdom; email: alexander.vanstone13@imperial.ac.uk

**Author Contributions**

AV, JCG, WB conceived the work. JCG, KDS fabricated the samples. AV performed experimental MOKE measurements and FMR measurements. AV performed analysis of FMR measurements. AV, JCG, KDS performed MFM measurements. AV wrote code for dipolar needle modelling and conversion to FMR spectra. DMA, TD wrote code for simulation of magnon spectra. AV drafted the manuscript, with contributions from all authors in editing and revision stages.

NOTES

The authors declare no competing financial interest.

ACKNOWLEDGMENTS

This work was supported by the Leverhulme Trust (RPG-2017-257) to WRB. TD and AV were supported by the EPSRC Centre for Doctoral Training in Advanced Characterisation of Materials (Grant No. EP/L015277/1). Simulations were performed on the Imperial College London Research Computing Service[43]. The authors would like to thank L. Cohen of Imperial College London and H. Kurebayashi of University College London for enlightening discussion and comments, and D. Mack for excellent laboratory management.